\documentstyle[12pt]{article}
\begin{document}

\begin{center}
{\small \bf ON THE ELECTRIC DIPOLE POLARIZABILITY \\
OF THE THREE-HADRON BOUND SYSTEM} \\[.4in]

\setlength{\baselineskip}{0.1in}
{\footnotesize V. F. KHARCHENKO}\footnote{Corresponding author.}
{\footnotesize and A. V. KHARCHENKO} \\[.05in]

{\footnotesize \it Bogolyubov Institute for Theoretical Physics,\\
National Academy of Sciences of Ukraine, UA - 03143, Kyiv, Ukraine\\
vkharchenko@bitp.kiev.ua} \\[.4in]
\end{center}

\begin{abstract}
\setlength{\baselineskip}{0.1in}
\noindent
{\footnotesize 
A simple analytical expression for the electric dipole polarizability of 
the three-hadron bound system having only one stable bound state has been 
derived neglecting by the higher orbital components of the off-shell 
three-body transition matrix at the energy of the bound state.
As a case in point, we have estimated the electric dipole polarizability 
of the triton, using a cluster triton wave function and the Hulth$\acute{e}$n 
potential to describe the related $p-n$ and $n-d$ bound states. }\\[.2in]
\end{abstract}

\vspace*{.1in}
\noindent {\bf 1. Introduction} \\ [.1in]
Data on the electric polarizabilities of the lightest nuclei $\alpha_E$ 
contain a valuable information on the nuclear force between nucleons. 

For the deuteron, the currently available values of $\alpha_E(^2\mbox{H})$  
obtained by the direct measurement of deviation from Rutherford
scattering of the deuteron on a heavy nucleus below the Coulomb barrier$^1$ 
and extracted from photoabsorption data$^2$,
\begin{equation}
\alpha_E(^2\mbox{H})=0.70 \pm 0.05\;\; \mbox{fm}^3\;(\mbox{Ref.
1})\;\;\mbox{and}\;\;\alpha_E(^2\mbox{H})=0.61 \pm 0.04\;\;
\mbox{fm}^3\;(\mbox{Ref. 2})\,,
\end{equation}
are a little distinguished between themselves.

For the nucleus $^3\mbox{He}$, contrastingly, the corresponding values of 
the polarizability, obtained by the direct way$^3$ and from experimental 
photoabsorption data$^4$,
\begin{equation}
\alpha_E(^3\mbox{He})=0.25 \pm 0.04\;\; \mbox{fm}^3\;(\mbox{Ref.
3})\;\;\mbox{and}\;\;\alpha_E(^3\mbox{He})=0.130 \pm 0.013\;\;
\mbox{fm}^3\;(\mbox{Ref. 4})\,,
\end{equation}
are distinguished by a factor of two.

For the nucleus $^3\mbox{H}$ no measurement of the 
electric dipole polarizability has been performed to the present time.   

The polarizability of the nucleus $^4\mbox{He}$ was experimentally 
found$^{4-6}$ to be less by about an order of magnitude than that of 
the deuteron.

Calculations of the deuteron electric polarizability, carried out
with the realistic nucleon-nucleon interaction potentials, lead to the
values of $\alpha_E(^2\mbox{H})=0.6328(17)\;\;\mbox{fm}^3$ (Ref.7), 
being closer to the data of Ref. 2 in (1). 

Furthermore, examining theoretically the anisotropy of the deuteron 
deformation in the electric field caused by the tensor character of 
the $n-p$ interaction, the separate components of the deuteron 
electric polarizability $\alpha^{|M|}_E$, the longitudinal component 
(with the deuteron spin along the electric field) $\alpha^1_E$ and 
the transverse one $\alpha^0_E$, have been calculated in Ref. 8.
(The above electric polarizability $\alpha_E(^2\mbox{H})$ is the
averaged value of the components $\alpha^{|M|}_E$,
$\alpha_E(^2\mbox{H})=\frac{2}{3}\alpha^1_E +
\frac{1}{3}\alpha^0_E$ .)

Computations of the scalar and tensor deuteron polarizabilities have
also been performed in the framework of the effective field theory
that uses space-time and global chiral symmetries of the quantum
chromodynamics consistently describing pion propagation and
relativistic effects$^{9-11}$. The results for the electric
deuteron polarizabilities obtained in the cited works
in the leading and next-to-leading orders agree with the values
calculated in the traditional nuclear physics with the application
of the potential models.

The results of calculations of the $^3\mbox{He}$ polarizability, 
$\alpha_E(^3\mbox{He})=0.145\;\; \mbox{fm}^3$ (Ref. 12), 
$\alpha_E(^3\mbox{He})=0.153(15)\;\; \mbox{fm}^3$ (Ref. 6) and 
$\alpha_E(^3\mbox{He})=0.149(5)\;\; \mbox{fm}^3$ (Ref. 13), support the 
experimental result in (2) that has been obtained from the photoabsorption 
data (Ref. 4). Also, the results of calculations of the triton polarizability, 
$\alpha_E(^3\mbox{H})=0.139(2)\;\; \mbox{fm}^3$ (Refs. 13 and 14), turned out 
to be close to those of the $^3\mbox{He}$ polarizability$^{6,12,13}$. 
(A little larger value of $\alpha_E(^3\mbox{He})$ may be assigned to the 
repulsive Coulomb interaction between the two protons in the nucleus 
$^3\mbox{He}$ causing the charge symmetry violation.) 

In the previous paper$^{15}$, we have predicted the value of the 
electric dipole polarizability of the only three-body lambda 
hypernucleus --- the lambda hypertriton $^3_{\Lambda}\mbox{H}$. 
It was found that $\alpha_E(^3_{\Lambda}\mbox{H})$ is close to 
3 $\mbox{fm}^3$ exceeding the polarizability of the ordinary 
three-body nuclei by an order of magnitude and even the recently 
measured polarizability of the nucleus $^6\mbox{He}$,
$\alpha_E(^6\mbox{He})=1.99(40)\;\; \mbox{fm}^3$ (Ref. 6).

This paper is devoted to development of a method of direct determination 
of the electric dipole polarizability of the three-particle bound system, 
leaning upon solution of the three-body problem at the negative bound-state 
energy without necessity of finding the continuum wave functions. In  Sec. 2,
a general formalism for determining the polarizability of the three-hadron 
complex is worked out. The electric polarizability of the three-hadron nucleus 
is expressed in terms of partial derivatives of the bound-state wave function 
in momentum space and selected higher partial components of the three-body 
off-shell transition matrix. In Sec. 3, the developed formalism is applied in
the case of a simple physically justified (cluster) model of the wave function 
of the triton (a bound complex of one proton and two neutrons). Neglecting 
higher partial components of the transition matrix we obtain a simple formula 
for estimation of the triton electric dipole polarizability. The results of 
corresponding calculations of $\alpha_E(^3\mbox{H})$ based on the known 
low-energy data for the $p-n$ and $d-n$ interactions are described in Sec. 4. 
Conclusions are drawn in Sec. 5. \\

\noindent {\bf 2. General formalism} \\ [.1in]
Previously, on the basis of the three-body formalism of the
effective interaction of a charged particle and a bound
complex$^{16-19}$, we have derived an expression for the
polarization potential of the two-hadron bound complex
that consists of charged and neutral hadrons (for example,
the deuteron)$^{17,20,21}$, starting immediately from the Faddeev
integral equations$^{22}$. In the case that the interaction
between the proton and the neutron composing the deuteron is central
and $S$-wave, the electric dipole polarizability of the deuteron is
given by$^{15}$
\begin{equation}
\alpha_E(^2\mbox{H})=\frac{2}{3}\frac{{e_p}^2}{\hbar^2 c^2}
\left(\frac{m_n}{m_{pn}}\right)^2
\int_{0}^{\infty} \frac{dk k^2}{2\pi^2} \frac{\mid \psi_d^\prime (k)\mid^2}
{\frac{k^2}{2\mu_{pn}}+B_d},
\end{equation}
where $e_p$ is the charge of the proton, $\mu_{pn}$ is the
proton-neutron reduced mass, $\mu_{pn}=m_pm_n/m_{pn}$,
$m_{pn}=m_p+m_n$ ($m_p$ and $m_n$ are the proton and neutron
masses), $\psi_d^\prime(k)\equiv d\psi_d(k)/dk$ is the first
derivative of the deuteron wave function in the momentum space in
the variable of the relative momentum $k$, and $B_d$
is the binding energy of the deuteron. The formula (3) is in
agreement with the expressions for $\alpha_E$ obtained in the case
of the separable S-wave pair potential in Refs. 17, 20 and 21.

According to the expression for the electric polarizability of the
two-particle bound complex (3), obtained assuming that the 
interaction in the $P$-wave orbital state is absent, the quantity 
$\alpha_E$ is completely determined by the wave function of the bound state 
of the complex. Hence, for the different interaction potentials producing 
identical bound-state wave functions (as an example, in the case of the 
two-body problem with the $S$-wave local Hulth\'{e}n interaction potential 
and with the $S$-wave separable potential having the Yukawa formfactor), even 
if distinct the corresponding scattering wave functions, the expressions 
for the polarizability $\alpha_E$ should be the same.

In particular, from this it follows that a closer determination of $\alpha_E$
is anticipated in the event of fitting the potential parameters to
bound-state data rather than scattering-state ones corresponding to higher
energies.

Here we obtain the expression for the electric dipole polarizability of
the three-hadron bound complex in the case when the system can
form, apart from the continuum, only one bound state.  We start from the
general expression for the polarizability of the complex , considering
the low-energy scattering of the three-body bound complex (with the binding
energy $B_0$) by the field of a particle 1 having the electric charge $e_1$.
For simplicity sake assume that the complex consists of one charged
particle 2 and two neutral particles (3 and 4). The initial kinetic energy of
the relative motion of the particle 1 and complex is taken to be far
lower than the breakup threshold energy of the complex. The effective
potential of the interaction between the charged particle 1 and complex
was found within the framework of the rigorous Watson-Feshbach
formalism$^{23,24}$ (for more details concerning application of the
above formalism for the three-body system see Refs. 17 - 19, 21 ). The electric
dipole polarizability $\alpha_E$ was determined as the strength of the
polarization potential at asymptotically large distances between the particle 1
and complex, ${\rho}_1$, greatly exceeding the size of the complex,
\begin{equation}
V_{pol}(\rho_1)=-\alpha_E\frac{e_1^2}{2\rho_1^4}\;,
\end{equation}
where  $\alpha_E$ is given as
\begin{equation}
\alpha_E=-2<\Psi_0\mid ({\bf D}_2\cdot\hat{\mbox{\boldmath$\rho$}}_1)
G^Q(-B_0) ({\bf D}_2\cdot\hat{\mbox{\boldmath$\rho$}}_1)\mid \Psi_0>\,,
\end{equation}
Here $\Psi_0$, ${\bf D}_2=e_2{\bf r}_2$, and $G^Q(-B_0)$ are the
wave function of the ground bound state of the three-hadron
complex corresponding to the binding energy $B_0$ (normalized to
unit, $<\Psi_0\mid\Psi_0>=1$), the operator of the dipole moment
of the charged particle 2 having the charge $e_2$, and the
truncated Green's operator of the complex $G^Q(E)=QG(E)$ at the
energy $E=-B_0$, respectively. The full Green's operator of the
complex $G(E)$ is given by $G(E)=(E-H_0-V)^{-1}$, where $H_0$ is
the kinetic energy operator and $V$ is the total interaction
potential, $V=v_{23}+v_{24}+v_{34}$, $v_{ij}$ is the potential of
the pair interaction between the particles $i$ and $j$ (the
potentials are assumed to be energy-independent), $Q=1-P$, $P$ is
the projection operator onto the complex ground state,
$P=\mid\Psi_0><\Psi_0\mid$. The quantity
${\mbox{\boldmath$\rho$}}_1$ in Eqs. (4) and (5) is the radius
vector specifying the relative position of the centre of mass of
the complex (composed of the particles 2, 3 and 4) with respect to
the charged particle 1, $\hat{{\mbox{\boldmath
$\rho$}}}_1\equiv{\mbox{\boldmath $\rho$}}_1/\rho_1$ is the unit
vector along ${\mbox{\boldmath $\rho$}}_1$, giving the direction
of the external electric field, created by the particle 1, ${\bf
r}_2$ is the radius vector of the charged constituent of the
complex, the particle 2, relative to the centre of mass of the
complex.

Expressing the full Green's operator $G(E)=(E-H_0-V)^{-1}$
through the transition matrix $T(E)$,
\begin{equation}
G(E)=G_0(E)+G_0(E)T(E)G_0(E),
\end{equation}
we write the operator $G^Q(E)$, which appears in the expression (5) for the
electric dipole polarizability of the three-body bound complex in the form
\begin{equation}
G^Q (E)=(1-P)[G_0(E)+G_0(E)T(E)G_0(E)]\;,
\end{equation}
where $G_0(E)=(E-H_0)^{-1}$ is the free Green's operator. The transition
matrix $T(E)$ is defined by the Lippmann-Schwinger integral equation
\begin{equation}
T(E)=V+VG_0(E)T(E)\;.
\end{equation}

In the case under consideration that the complex has only one bound state
with the energy $E=-B_0$, the transition matrix may be written as the
sum of the pole, $\check{T}(E)$, and smooth, $\tilde{T}(E)$, parts,
\begin{equation}
T(E)=\check{T}(E)+\tilde{T}(E)\;,
\end{equation}
where
\begin{equation}
\check{T}(E)=\frac{\mid\Gamma_0\rangle\langle\Gamma_0\mid}{E+B_0}\;,
\end{equation}
\begin{equation}
\mid\Gamma_0\rangle=G_0^{-1}(-B_0)\mid\Psi_0\rangle=V\mid\Psi_0\rangle\;.
\end{equation}
At the point $E=-B_0$ the operator $\tilde{T}(E)$ may be shown to have
the form
\begin{equation}
\tilde{T}(-B_0)=-c_1\mid\Gamma_0\rangle\langle\Gamma_0\mid +\cdots\;,
\end{equation}
with
\begin{equation}
c_1=\langle\Psi_0\mid G_0(-B_0)\mid\Psi_0\rangle\;.
\end{equation}

The factored term in (12) is the smooth part of the dominant partial
component of the three-body transition matrix $T(E)$ with zero
relative orbital momenta. The additional terms in (12) indicated by
the ellipsis contain the higher orbital components of $T(E)$ at $E=-B_0$,
which are wholly smooth functions of $E$, however, being less important,
they are disregarded here.

In view of (7) and (9), the operator $G^Q(E)$ may be written as
\begin{equation}
G^Q(E)=G_0^Q(E)+\check{G}^Q(E)+\tilde{G}^Q(E)
\end{equation}
while the corresponding terms are given by
\begin{equation}
\begin{array}{ccl}
G_0^Q(E) & = & (1-P)G_0(E), \\
\check{G}^Q(E) & = & (1-P)G_0(E)\check{T}(E)G_0(E), \\
\tilde{G}^Q(E) & = & (1-P)G_0(E)\tilde{T}(E)G_0(E).
\end{array}
\end{equation}

The value of the matrix element
\begin{equation}
\begin{array}{ccc}
\langle\Psi_0\mid ({\bf D_2}\cdot\hat{\mbox{\boldmath$\rho$}}_1) G^Q(E)
({\bf D_2}\cdot\hat{\mbox{\boldmath$\rho$}}_1)\mid\Psi_0\rangle & = &
\langle\Psi_0\mid({\bf D_2}\cdot\hat{\mbox{\boldmath$\rho$}}_1) G_0^Q(E)
({\bf D_2}\cdot\hat{\mbox{\boldmath$\rho$}}_1)\mid\Psi_0\rangle \\
& + & \langle\Psi_0\mid ({\bf
D_2}\cdot\hat{\mbox{\boldmath$\rho$}}_1)\check{G}^Q(E) ({\bf
D_2}\cdot\hat{\mbox{\boldmath$\rho$}}_1)\mid\Psi_0\rangle \\
& + & \langle\Psi_0\mid({\bf D_2}\cdot\hat{\mbox{\boldmath$\rho$}}_1)
\tilde{G}^Q(E)({\bf D_2}\cdot\hat{\mbox{\boldmath$\rho$}}_1)\mid\Psi_0\rangle
\end{array}
\end{equation}
appearing in the expression (5) at the energy $E=-B_0$ is obtained with the
use of the relations (10)---(12) in the second and third terms on the
right-hand side of Eq. (16) and with evaluating the indeterminate form of the
type 0/0 in the second term. Furthermore, it is easy to verify
that the third term vanishes at the point $E=-B_0$.

As a result, the expression for the electric dipole polarizability (5)
becomes then
\begin{equation}
\begin{array}{rcl}
\alpha_E & = & -\;\;2\{\langle\Psi_0\mid ({\bf
D_2}\cdot\hat{\mbox{\boldmath$\rho$}}_1) G_0(-B_0)({\bf
D_2}\cdot\hat{\mbox{\boldmath$\rho$}}_1)\mid\Psi_0\rangle \\
\; &  & +\;\;\langle\Psi_0\mid
({\bf D_2}\cdot\hat{\mbox{\boldmath$\rho$}}_1) \mid\Psi_0\rangle
\cdot [ c_1 \langle\Psi_0\mid ({\bf
D_2}\cdot\hat{\mbox{\boldmath$\rho$}}_1) \mid\Psi_0\rangle \\
\; &   & -\;\;\langle\Psi_0\mid ({\bf D_2}\cdot\hat{\mbox{\boldmath$\rho$}}_1)
G_0(-B_0)\mid\Psi_0\rangle - \langle\Psi_0\mid G_0(-B_0)({\bf
D_2}\cdot\hat{\mbox{\boldmath$\rho$}}_1) \mid \Psi_0 \rangle ] \} .
\end{array}
\end{equation}

Further simplification of the expression (17) for $\alpha_E$ occurs
if the wave function of the bound complex $\Psi_0$ is
characterized by a definite parity (as a consequence of the spatial
reflection invariance of the interaction potential $V$). In such a case
the matrix elements
$\langle\Psi_0\mid ({\bf D_2}\cdot\hat{\mbox{\boldmath$\rho$}}_1)
\mid\Psi_0\rangle\;,
\langle\Psi_0\mid G_0(-B_0)({\bf D_2}\cdot\hat{\mbox{\boldmath$\rho$}}_1)
\mid\Psi_0\rangle \mbox{ and } \\
\langle\Psi_0\mid ({\bf D_2}\cdot\hat{\mbox{\boldmath$\rho$}}_1)
G_0(-B_0)\mid\Psi_0\rangle$
in (17) vanish due to integration over the angular variables. The
expression (17) then reduces to the simple form
\begin{equation}
\alpha_E=
-2\langle\Psi_0\mid({\bf D_2}\cdot\hat{\mbox{\boldmath$\rho$}}_1) G_0(-B_0)
({\bf D_2}\cdot\hat{\mbox{\boldmath$\rho$}}_1)\mid\Psi_0\rangle
\end{equation}
containing only the free Green's operator at the negative energy $E=-B_0$.

The bound complex being considered below is a three-hadron nucleus composed
of the proton $p$ (the charged particle 2), the neutron $n$ (the neutral
particle 3) and the neutral hadron $h$ (the particle 4). 
The symbol $h$ stands for $n$ (the neutron) in the case of the
triton and for $\Lambda$ (the lambda hyperon) in the case of the lambda 
hypertriton.

In the momentum representation the operators of the dipole moment ${\bf D}_2$
and the free propagator $G_0(-B_0)$ in the Eq. (18) are given by
\begin{equation}
\begin{array}{c}
{\bf D}_2 = ie_p \left( \frac{m_n}{m_{pn}}{\bf \nabla}_{\bf k} +
\frac{m_h}{m_{pnh}}{\bf \nabla}_{\bf p} \right) \;, \\
\langle{\bf kp}\mid G_0(-B_0)\mid {\bf k}^{\prime} {\bf p}^{\prime}
\rangle= -(2\pi)^6 \delta ({\bf k}-{\bf k}^{\prime}) \delta
({\bf p}-{\bf p}^{\prime}) \left( B_0+\frac{k^2}{2\mu_{pn}}+
\frac{p^2}{2\mu_{pn,h}}\right) ^{-1}\;.
\end{array}
\end{equation}
where ${\bf k}$ and ${\bf p}$ are the Jacobi momentum variables describing the
relative motion of particles $p$ and $n$ and that of the particle $h$ with
respect to the centre of mass of ($p$,$n$),
\begin{equation}
{\bf k}=\frac{m_n {\bf k}_p-m_p {\bf k}_n}{m_{pn}},\;\; {\bf
p}=\frac{m_h({\bf k}_p+{\bf k}_n)-m_{pn}{\bf k}_h}
{m_{pnh}}\;.
\end{equation}
Here ${\bf k}_i$ is the momentum of the particle $i$, $\mu_{pn,h}$ is the
reduced mass of the system ($p,n$) with the mass $m_{pn}=m_p+m_n$ and
the hyperon $h$ with the mass $m_h$,
$\mu_{pn,h}=m_{pn}m_h/m_{pnh}$,
$m_{pnh}= m_p+m_n+m_h$, the binding energy of the three-hadron bound complex
$B_0$ is equal to the sum of the deuteron binding energy
$B_d=\kappa_d^2/{2\mu_{pn}}$ and the separation energy of the hyperon $h$,
$B_h=\kappa_{h}^2/{2\mu_{d,h}}$, $B_0=B_d+B_{h}$,
$\mu_{d,h}=m_d m_{h}/m_{dh}$, $m_d$ is the
deuteron mass, $m_{dh}=m_d+m_h$, $\Psi_0({\bf k},{\bf p})$ is
the normalized wave function of the three-hadron nucleus in momentum space,
${\bf \nabla}_{\bf k}\equiv\frac{\partial}{\partial{\bf k}}$ is the gradient
operator.

When the full wave function $\Psi_0$ in the general formula (18) has the form
of the product of spatial and spin functions, we may omit spin variables from
consideration. In the momentum space, substituting the expressions (19) into
(18), the formula for the electric dipole polarizability of the three-hadron
bound system may be written as
\begin{equation}
\alpha_E(pnh)=2\frac{e_p^2}{\hbar^2 c^2}
\int \frac{d{\bf k}d{\bf p}}{(2\pi)^6} \frac{\mid \langle {\bf k}{\bf p}\mid
\hat{\mbox{\boldmath$\rho$}}_1\cdot\left( \frac{m_n}{m_{pn}} {\bf \nabla}_{\bf
k}+\frac{m_{h}}{m_{pnh}} {\bf \nabla}_{\bf p}\right) \Psi_{pnh}\rangle \mid ^2}
{\frac{k^2}{2\mu_{pn}}+\frac{p^2}{2\mu_{pn,h}}+B_0},
\end{equation}
where $\Psi_{pnh}$ is the spatial wave function of the three-hadron nucleus.

We are reminded that the formula for the electric polarizability of
the three-hadron bound system (21) has been derived assuming that the higher
orbital components of the off-shell three-body transition matrix at the
negative energy $E=-B_0$ are negligibly small. Similar to the expression (3)
for the electric dipole polarizability of the two-body complex, the obtained
expression for the polarizability of the three-body complex is essentially
determined by the first derivatives of its bound-state wave function with
respect to the Jacobi momentum variables.

Below, using a simple model wave function we employ our formula (21) to
evaluate the electric dipole polarizability of the simplest three-body
nucleus, the triton $^3\mbox{H}$ containing only one charged particle 
(the proton).

It is worth noting that the parametrized analytical form for the Faddeev
triton wave function generated with the Reid soft-core potential in Ref. 25
is quite suitable to use in the expression of the type (21) with the aim
to calculate the triton electric polarizability of the triton.\\

\vspace*{.1in}
\noindent {\bf 3. Model wave function of the triton} \\ [.1in]  
In this article we study the electric dipole polarizability of the triton 
nucleus containing the proton (the particle 2) and two neutrons (the 
particles 3 and 4). To estimate the polarizability, we use the clustered 
(deuteron + neutron) triton wave function that must be antisymmetrized 
with respect to the identical fermions.

For simplicity sake, without introducing the isobaric formalism, we have to
antisymmetrize only with respect to the two neutrons. This antisymmetrization
naturally occurs, if we shall restrict our consideration only the dominant
symmetric part of the space wave function of the triton, since its
spin wave function $\chi_t \equiv \chi_{SM_S}$ (with $S=1/2$) is already
antisymmetric relative to the permutation of the neutrons,
\begin{equation}
\chi_{\scriptstyle \frac{1}{2} \frac{1}{2}} (3,4;2)=\frac{1}{\sqrt{2}}
\left( \alpha_3\beta_4-\alpha_4\beta_3 \right)\alpha_2.
\end{equation}
where $\alpha_i$ and $\beta_i$ are the spin functions of the particle $i$ 
with the spin projections +1/2 and -1/2.

Thus, in the framework of the cluster model of the triton as a $n+d$ system,
the space part of the triton wave function must be symmetric in the
interchange of the two neutrons, 
\begin{equation}
\Psi_t({\bf k},{\bf p})=\frac{N_n}{\sqrt{2}}\left[\psi_d(k_{23})\phi_n(p_4)+
\psi_d(k_{24})\phi_n(p_3)\right],
\end{equation}
where $\phi_n(p)$ is the normalized bound-state wave function describing
relative motion of the neutron and centre of mass of the deuteron,
the Jacobi momentum variables ${\bf k}_{23}\equiv{\bf k}$ and
${\bf p}_4\equiv{\bf p}$ are defined by the relations (20),
\begin{equation}
{\bf k}_{24}=\frac{m_n}{m_{pn}}{\bf k}+\frac{m_p m_{pnn}}{{m_{pn}}^2}
{\bf p},\;\; {\bf p}_3={\bf k}-\frac{m_n}{m_{pn}}{\bf p}\;,
\end{equation}
and the coefficient $N_n$ ensures normalization of the whole
triton wave function to 1,
$$
N_n=\left(1+\langle\psi_d(k)\phi_n(p)+\psi_d(k_{24})\phi_n(p_3)\rangle
\right)^{-1/2}\;.
$$

Substituting the model wave function (23) into the formula (21),
we get the following expressions for the electric dipole polarizabilities
of the triton, 
\begin{equation}
\alpha_E(^3\mbox{H})=\alpha_E^0(pnn)+\alpha_E^{ex}(pnn)
\end{equation}
with
\begin{eqnarray}
\alpha_E^0(pnn)=\frac{4}{3}\frac{{e_p}^2}{\hbar^2}\frac{m_pm_n}{m_{pn}}
N^2_n\int_{0}^{\infty}\frac{p^2dp}{2\pi^2}\left\{ \left(\frac{m_n}{m_{pnn}}
\right)^2 I_n(p)\left[\frac{d\phi_n(p)}{dp}\right]^2\right. \nonumber \\ [1mm]
+\left.\left( \frac{m_n}{m_{pn}}\right)^2 J_n(p)
\left[\phi_n(p)\right]^2 \right\} \label{line2}
\end{eqnarray}
and
\begin{eqnarray}
\alpha_E^{ex}(pnn)=4\frac{{e_p}^2}{\hbar^2}\frac{m_pm_n}{m_{pn}} N^2_n
\int\int\frac{d{\bf k}d{\bf p}}{(2\pi)^6}\frac{1}{k^2+[C_n(p)]^2} \nonumber \\
\cdot \left\{ \left(\frac{m_n}{m_{pnn}}\right)^2
\psi_d(k)\psi_d(k_{24})\frac{d\phi_n(p)}{dp}\frac{d\phi_n(p_3)}{dp_3}
(\hat{\mbox{\boldmath$\rho$}}_1\cdot\hat{p})
(\hat{\mbox{\boldmath$\rho$}}_1\cdot\hat{p}_3)\right. \nonumber \\
+\left.\left(\frac{m_n}{m_{pn}}\right)^2
\frac{d\psi_d(k)}{dk}\frac{d\psi_d(k_{24})}{dk_{24}}\phi_n(p)\phi_n(p_3)
(\hat{\mbox{\boldmath$\rho$}}_1\cdot\hat{k})
(\hat{\mbox{\boldmath$\rho$}}_1\cdot\hat{k}_{24})\right\}\;. \label{line3}
\end{eqnarray}
Here the hat signifies a unit vector, the functions $I_n(p)$ and $J_n(p)$
are defined by
\begin{eqnarray}
I_n(p)=\int_{0}^{\infty}\frac{dk k^2}{2\pi^2} \frac{[\psi_d(k)]^2}
{{k^2}+[C_n(p)]^2}\;,\quad J_n(p)=\int_{0}^{\infty} \frac{dk k^2}{2\pi^2}
\frac{[d\psi_d(k)/dk]^2}{{k^2}+[C_n(p)]^2}\;, \label{line1}\\ [2mm]
{}[C_n(p)]^2 \equiv \frac{m_pm_n}{m_{pn}m_n} \left\{
\frac{m_{pnh}}{m_{pn}} p^2 +\frac{m_{dh}}{m_d} \kappa_h^2
\right\} +\kappa_d^2\;\;. \nonumber
\end{eqnarray}

The wave functions $\psi_d(k)$ and $\phi_n(p)$ in
the equations (26)--(28), which determine the model wave function of the
triton, were found by solving analytically the corresponding two-body problems.  
Note that the $p-n$ interaction is relatively weak resulting to rather small 
deuteron binding energy. This interaction can be well described using even 
a simple separable rank-1 potential. Another situation exists in the case of 
the $n-d$ interaction that is stronger than the $p-n$ interaction.
In addition to the $n-d$ bound ground state (the triton), the $n-d$ interaction
supports also one virtual state (that is reflected in a rather small value
of the $n-d$ scattering length). In this connection the $n-d$ interaction
can be described by a separable potential of the rank no less than 2 or
by a local potential.

Here, in the case of the triton, both the wave functions $\psi_d(k)$ and
$\phi_n(p)$ were determined by solving the corresponding two-body bound-state
problems with the local Hulth$\acute{e}$n (H) potential$^{26}$,
\begin{equation}
v^H(r)=-v_0\left[\exp(qr)-1 \right]^{-1}\;,
\end{equation}
employed to describe the proton-neutron interaction, $v^H_{pn}(r)$, and the
effective interaction between the neutron and the deuteron (as a structureless
object), $v^H_n(\rho)$. (The radius vector variables ${\bf r}$ and
\mbox{\boldmath $\rho$} in configuration space correspond to the variables
${\bf k}$ and ${\bf p}$ in momentum space.) \\

\vspace*{.1in}
\noindent {\bf 4. Calculations and discussion of results} \\ [.1in]
The electric dipole polarizability of the triton $\alpha_E(^3\mbox{H})$ was 
calculated by the formulae (25) -- (28).

The parameters of the  $p-n$ interaction potentials were fitted to
the experimental values of the deuteron binding energy $B_d$
and the triplet $p-n$ scattering length $^3a_{pn}$,
\begin{equation}
B_d=2.224575(9)\;\mbox{MeV}\;\; \mbox{(Ref. 27)}\,,\;\;
^3a_{pn}=5.424(3)\;\mbox{fm}\;\; \mbox{(Ref. 28)}.
\end{equation}

The parameters of the effective $n-d$ interaction potential (29) were
determined using the experimental values of the separation energy of the
neutron (needed to remove one neutron from the triton, $B_n=B_t-B_d$) and
the doublet $n-d$ scattering length $^2a_{nd}$,
\begin{equation}
B_t=8.481855(13)\;\mbox{MeV}\;\; \mbox{(Ref. 29)}\,,\;\;
^2a_{nd}=0.65(4)\;\mbox{fm}\;\; \mbox{(Ref. 30)}.
\end{equation}

The fitted values of the parameters ($\gamma=v_0q^{-3}$ and $\beta=q+\kappa$) 
of the local Hulth$\acute{e}$n $p-n$ and $n-d$ ineraction potentials (29), 
$v_{pn}$ and $v_n$,were found to be 
\begin{equation}
2\mu_{pn}\gamma_{pn}=1.3184\;\mbox{fm}\,,\;\;\beta_{pn}=1.3146\;\mbox{fm}^{-1}\,;
2\mu_{nd}\gamma_n=5.4689\;\mbox{fm}\,,\;\;\beta_n=0.9552\;\mbox{fm}^{-1}\,.
\end{equation}

For the deuteron, the electric dipole polarizability obtained from Eq.(3)
for the local Hulth$\acute{e}$n potential with the parameters (32) takes the value
$\alpha_E^{\mbox{{\scriptsize H}}}(^2\mbox{H})=0.6442\;\mbox{fm}^3$.
It is known $^{7,31}$ that the S-wave asymptotic normalization constant
$A_S(^2\mbox{H})$ accounts for most of the polarizability
$\alpha_E(^2\mbox{H})$ having regard to a rather high probability of
that the slightly bound nucleons in the deuteron are at distances outside
of the range of the nuclear force. If both the parameters of the potential are
fitted to only the data that concerns to the bound $p-n$ state, for example,
with the use of the binding energy $B_d$ (30) and
\begin{equation}
A_S(^2\mbox{H})=0.8845(8)\;\; \mbox{fm}^{-1/2}\;\;\mbox{(Ref. 32)}\;,
\end{equation}
we find for the deuteron polarizability the value
\begin{equation}
\alpha_E^{\mbox{{\scriptsize H}}}(^2\mbox{H})=0.6292\;\mbox{fm}^3\,.
\end{equation}
With the use of the tensor separable potential that corresponds to $A_S=0.8843\;\;
\mbox{fm}^{-1/2}$ the deuteron polarizability increases still further --- to 
the value $\alpha_E=0.6311\;\; \mbox{fm}^3$ (Ref. 8) --- approaching the values 
obtained with the realistic potentials, $\alpha_E=0.6328(17)\;\; \mbox{fm}^3$ 
(Ref. 7).

In this paper the electric dipole polarizability of the triton given by the
formulae (25) -- (28) has been calculated with the use of
the local Hulth$\acute{e}$n potential that allows of finding the wave functions
$\psi_d(k)$ and $\phi_n(p)$ in the analytical form. Fitting the parameters
of the potential to the data for $B_d$ and $^3a_{pn}$ of the $p-n$ system
(30) and for $B_n$ and $^2a_{nd}$ of the $n-d$ system, (30) and (31), we have
obtained for the triton polarizability the value
\begin{equation}
\alpha_E^{\mbox{{\scriptsize H}}}(^3\mbox{H})=0.235\;\mbox{fm}^3\;.
\end{equation}
In this case, the direct and exchange terms in (25) are found to contribute
to the triton polarizability almost equally:
$$
\alpha_E^0(pnn)=0.109\;\mbox{fm}^3\;,\;\;
\alpha_E^{ex}(pnn)=0.126\;\mbox{fm}^3\;.
$$

If the parameters of the $p-n$ interaction potential are fitted to the data
characterizing only $p-n$ bound state (the deuteron), $B_d$ and $A_S(^2\mbox{H})$ 
(the $S$-wave asymptotic normalization), we obtain a slightly less value of the
triton polarizability, 
$\alpha_E^{\mbox{{\scriptsize H}}}(^3\mbox{H})=0.225\;\mbox{fm}^3\;.$ The
decrease of the polarizability in this case occurs due to the decrease of
the quantity $A_S$ (the experimental value of $A_S$ (33) is less than the
value of $A_S$ for the Hulth$\acute{e}$n wave function corresponding to
the potential fitted to the data $B_d$ and $^3a_{pn}$
($A_S(^2\mbox{H})=0.8960\;\;\mbox{fm}^{-1/2}$)).

Unfortunately, the value $\alpha_E(^3\mbox{H})$ has not been measured yet.
Although our tentative estimate of the triton polarizability, carried out 
on the basis of the direct calculation by the formula (21) but using the 
cluster wave function, leads to the result for $\alpha_E(^3\mbox{H})$ that 
is consistent with one of two data for the polarizability of the mirror nucleus 
$^3\mbox{He}$ (Ref. 1), a new calculation of $\alpha_E(^3\mbox{H})$ with 
the use of a more reasonable triton wave function in (21) would be valuable.\\

\vspace*{.1in}
\noindent {\bf 5. Summary and conclusions} \\[.1in]
Leaning upon on the analytical structure of the three-body
transition matrix, a consistent formalism for determination of the
electric dipole polarizability of a three-hadron bound complex
consisting of one charged and two neutral particles and having
only one stable bound state has been worked out. A simple
expression for the electric dipole polarizability of
the three-hadron bound system has been derived assuming
that the higher orbital components of the three-body off-shell 
transition matrix at the negative energy $E=-B_0$ are negligibly
small. In this case, the polarizability is expressed in terms of 
the first partial derivatives of the bound-state wave function with 
respect to the Jacobi momentum variables of the complex (Eq. (21)). 

Applying the cluster model for the triton wave function and the
Hulth$\acute{e}$n interaction potential to describe the $p-n$ and $n-d$
bound systems, we have obtained for the electric polarizability of the
triton the approximate estimate
$\alpha_E^{\mbox{{\scriptsize H}}}(^3\mbox{H})=0.23\;\mbox{fm}^3 $(using
the low energy data for $B_d$ and $^3a_{pn}$ of the $p-n$ system and for 
$B_n$ and $^2a_{nd}$ of the $n-d$ system). Under conditions 
that there is presently no direct measurement of the quantity
$\alpha_E(^3\mbox{H})$ and the results of the experiments 
for the polarizability of the mirror nucleus $^3\mbox{He}$ 
(Refs. 3 and 4), are inconsistent, more complicated calculations of 
the triton polarizability on the basis of the expression (21) and modern 
wave functions for the $^3\mbox{H}$ bound state are worth to be performed. \\

\vspace*{.1in}
\noindent {\footnotesize {\bf References}
\vspace*{.1in}
\begin{itemize}
\setlength{\baselineskip}{.1in}
\item[{\tt 1.}]N. L. Rodning, L. D.Knutson, W. G. Lynch, and M.
           B. Tsang, {\it Phys. Rev. Lett.} {\bf 49}, 909 (1982).
\item[{\tt 2.}]J. L. Friar, S. Fallieros, E. L. Tomusiak, D. Skopik and
           E. G. Fuller, {\it Phys. Rev.} {\bf C27}, 1364 (1983).
\item[{\tt 3.}]F. Goeckner, L. O. Lamm and L. D. Knutson,
           {\it Phys. Rev.} {\bf C43}, 66 (1991).
\item[{\tt 4.}]G. A. Rinker, {\it Phys. Rev.} {\bf A14}, 18 (1976).
\item[{\tt 5.}]J. L. Friar, {\it Phys. Rev.} {\bf C16}, 1540 (1977).
\item[{\tt 6.}]K. Pachucki and A. M. Moro, {\it Phys. Rev.} {\bf A75}, 032521
           (2007).
\item[{\tt 7.}]J. L. Friar and G. L. Payne, {\it Phys. Rev.} {\bf C55}, 2764
           (1997).
\item[{\tt 8.}]A. V. Kharchenko, {\it Nucl. Phys.} {\bf A617}, 34 (1997).
\item[{\tt 9.}]J.-W. Chen, H. W. Grie{\ss}hammer, M. J. Savage and
           R. P. Springer, {\it Nucl. Phys.} {\bf A644}, 221 (1998);
           nucl-th/9806080.
\item[{\tt 10.}]D. R. Phillips, G. Rupak and M. J. Savage, {\it Phys. Lett.}
           {\bf B473}, 209 (2000).
\item[{\tt 11.}]X. Ji and Y. Li, {\it Phys. Lett.} {\bf B591}, 76 (2004).
\item[{\tt 12.}]W. Leidemann, in {\it Few-Body Problems in Physics '02}, ed. 
           R.Krivec, B. Golli, M. Rosina and S. Sirca (Springer Verlag, Wien, 
           New York, 2003), vol. 14, p. 313.
\item[{\tt 13.}]I. Stetcu, S. Quaglioni, J. L. Friar, A. C. Hayes and 
           Petr Navr$\acute{a}$til, {\it Phys. Rev.} {\bf C79}, 064001 (2009).
\item[{\tt 14.}]V. D. Efros, W. Leidemann and G. Orlandini, {\it
           Phys. Lett.} {\bf B408}, 1 (1997).
\item[{\tt 15.}]V. F. Kharchenko and A. V. Kharchenko, {\it Collected Physical
              Papers} {\bf 7}, 432 (2008); nucl-th/0811.2565.
\item[{\tt 16.}]V. F. Kharchenko and S. A. Shadchin, {\it Few-Body Systems}
           {\bf 6}, 45 (1989); {\it Yad. Fiz.} {\bf 45}, 333 (1987).
\item[{\tt 17.}]V. F. Kharchenko and S. A. Shadchin S.A., {\it Three-body
           theory of the effective interaction between a particle and a
              two-particle bound system}, preprint ITP-93-24E (Institute for
           Theoretical Physics, Kyiv, 1993).
\item[{\tt 18.}]V. F. Kharchenko and S. A. Shadchin, {\it Ukrainian J. Phys.}
           {\bf 42}, 11 (1997).
\item[{\tt 19.}]V. F. Kharchenko, {\it J. Phys. Studies} {\bf 4},
           245 (2000).
\item[{\tt 20.}]V. F. Kharchenko, S. A. Shadchin and S. A. Permyakov, {\it
           Phys. Lett.} {\bf B199}, 1 (1987).
\item[{\tt 21.}]V. F. Kharchenko and S. A. Shadchin, {\it Ukrainian J. Phys.}
           {\bf 42}, 912 (1997).
\item[{\tt 22.}]L. D. Faddeev, {\it Zh. Eksp. Teor. Fiz.} {\bf 39}, 1459 (1960).
\item[{\tt 23.}]N. C. Francis and K. M. Watson, {\it Phys. Rev.} {\bf 92},
           291 (1953).
\item[{\tt 24.}]H. Feshbach, {\it Ann. Phys. (N.Y.)} {\bf 5}, 357 (1958);
           {\bf 19}, 287 (1962).
\item[{\tt 25.}]Ch. Hajduk, A. M. Green and M. E. Sainio, {\it Nucl. Phys.}
           {\bf A337}, 13 (1980).
\item[{\tt 26.}]L. Hulth$\acute{e}$n and M. Sugawara, in {\it Encyclopedia of
           Physics}. Vol. {\bf 39}, ed. S. Fl$\ddot{u}$gge (Springer, Berlin,
              1957).
\item[{\tt 27.}]C. Van der Leun and C. Alderliesten, {\it Nucl. Phys.}
           {\bf A380}, 261 (1982).
\item[{\tt 28.}]L. Koester and W. Nistler, {\it Z. Phys.} {\bf A272}, 189
           (1975).
\item[{\tt 29.}]A. H. Wapstra and G. Audi, {\it Nucl.  Phys.} {\bf A432}, 1
           (1985).
\item[{\tt 30.}]W. Dilg, L. Koester and W. Nistler, {\it Phys. Lett.} {\bf
           B36}, 208 (1971).
\item[{\tt 31.}]J. L. Friar and S. Fallieros,{\it Phys. Rev.} {\bf C29},
           232 (1984).
\item[{\tt 32.}]J. J. de Swart, C. P. F. Terheggen and V. G. J. Stoks,
           {\it Proc. of the Third Int. Symposium "Dubna Deuteron 95"},
           Dubna, Russia, 1995; nucl-th/9509032; J. J. de Swart, R. A. M.
           Klomp, M. C. M. Rentmeester and Th. A. Rijken,
           {\it Few-Body Syst. Suppl.} {\bf 99} (1995).

\end{itemize}}

\end{document}